\documentstyle[12pt,epsfig]{article}
\def\lsim{\raise0.3ex\hbox{$<$\kern-0.75em\raise-1.1ex\hbox{$\sim$}}}
\def\gsim{\raise0.3ex\hbox{$>$\kern-0.75em\raise-1.1ex\hbox{$\sim$}}}

\def\pom{{I\!\!P}}

\def\beq{\begin{equation}}
\def\eeq{\end{equation}}
\def\bea{\begin{eqnarray}}
\def\eea{\end{eqnarray}}
\def\bq{\begin{quote}}
\def\eq{\end{quote}}

\newcommand{\rr}{\mbox{$r$}}
\def\gappeq{\mathrel{\rlap {\raise.5ex\hbox{$>$}}
{\lower.5ex\hbox{$\sim$}}}}

\def\lappeq{\mathrel{\rlap{\raise.5ex\hbox{$<$}}
{\lower.5ex\hbox{$\sim$}}}}

\def\Toprel#1\over#2{\mathrel{\mathop{#2}\limits^{#1}}}

\newcommand{\rk}{\mbox{\boldmath $k$}}

\def\pom{{I\!\!P}}

\begin{document}
\pagestyle{empty}
\begin{center}
{\bf DIPOLE MODEL FOR DOUBLE MESON PRODUCTION IN TWO-PHOTON INTERACTIONS AT HIGH ENERGIES}
\\

\vspace*{1cm}
 V.P. Gon\c{c}alves $^{1}$, M.V.T. Machado  $^{2,\,3}$\\
\vspace{0.3cm}
{$^{1}$ Instituto de F\'{\i}sica e Matem\'atica,  Universidade
Federal de Pelotas\\
Caixa Postal 354, CEP 96010-090, Pelotas, RS, Brazil\\
$^{2}$ \rm Universidade Estadual do Rio Grande do Sul - UERGS\\
 Unidade de Bento Gon\c{c}alves. CEP 95700-000. Bento Gon\c{c}alves, RS, Brazil\\
$^{3}$ \rm High Energy Physics Phenomenology Group, GFPAE  IF-UFRGS \\
Caixa Postal 15051, CEP 91501-970, Porto Alegre, RS, Brazil}\\
\vspace*{1cm}
{\bf ABSTRACT}
\end{center}

\vspace*{1.5cm} \noindent

\vspace*{1.3cm} \noindent \rule[.1in]{17cm}{.002in}

\vspace{-3.5cm} \setcounter{page}{1} \pagestyle{plain}


In this work the double vector meson  production in two-photon interactions at high energies  is investigated considering saturation physics. 
We extend the color dipole picture for this process and study the energy and virtuality dependence of the forward differential cross section. Comparison with previous results is presented and the contribution of the different photon polarizations is estimated.

\vspace{0.5cm}

\section{Introduction}

The high energy limit of the perturbative QCD is characterized by a center-of-mass energy which is much larger than the  hard scales present in the problem. In  this regime the parton densities inside the projectiles grow as energy increases, leading to the rise of the cross sections. As long the energy is not too high, we have low values of the partonic density and  the QCD dynamics is described by  linear (BFKL/DGLAP)  evolution equations \cite{bfkl,dglap}. However, at higher energies the parton density increases and the scattering amplitude tends to   the unitarity limit. Thus, a linear description  breaks down and one enters the saturation regime, where the dynamics is described by a nonlinear evolution equation and the parton densities saturate \cite{BK,CGC}.  The transition line between the linear and nonlinear regimes is characterized by the saturation scale $Q_{\mathrm{sat}} (x)$, which is energy dependent and  sets the
critical transverse size for the unitarization of the cross
sections. In other words, unitarity is restored by including
nonlinear corrections in the evolution equations.
Such effects are small for $k_{\perp}^2 > Q_{\mathrm{sat}}^2$ and very strong for $k_{\perp}^2
< Q_{\mathrm{sat}}^2$, leading to the saturation of the scattering amplitude, where $k_{\perp}$ is the typical hard scale present in the process.
 The successful description of all
inclusive and diffractive deep inelastic data at the collider HERA
by saturation models \cite{GBW,bgbk,kowtea,iancu_munier}  suggests that these effects might become important
in the energy regime probed by current colliders. 
Furthermore, the saturation model was  extended to two-photon interactions at high energies in Ref. \cite{Kwien_Motyka}, also providing a very good description of the data on the $\gamma \gamma$ total cross section, on the photon structure function at low $x$ and on the $\gamma^* \gamma^*$ cross section.
The formalism used in Ref. \cite{Kwien_Motyka} is based on the dipole picture \cite{dipole}, with the  $\gamma^* \gamma^*$ total cross sections 
being described  by the interaction of two color dipoles, in which the virtual photons fluctuate into (For previous analysis using the dipole picture see, e.g.,  Refs. \cite{nik_photon,dona_dosch}). The dipole-dipole cross section is modeled considering the saturation physics.
The successful descriptions of the $\gamma \gamma$ interactions and  light/heavy vector meson production for $ep$ collisions at HERA are our main motivations to extend this formalism to describe the double meson production and analyze the effects of the saturation physics.


In the last few years the double  meson production has been studied considering different approaches and approximations for the QCD dynamics \cite{dona_dosch,motyka,motyka_ziaja,per1,double_meson,vic_sauter,wallon,Enberg,DIvanov}. In particular, in our previous paper in Ref. \cite{per1}, we have performed a phenomenological analysis for the double $J/\Psi$ production using the forward LLA BFKL solution. In that case, the hard scale was set by the charm quark mass. There, we also studied the possible effects of corrections at next to leading approximation (NLA) level to the BFKL kernel investigating the influence of a smaller  effective hard Pomeron intercept. Afterwards, in Ref. \cite{vic_sauter} the non-forward solution was considered for a larger set of possible vector meson pairs, where the large $t$ values provide the perturbative scale. Moreover, in that paper the double vector meson production in real photon interactions was studied, the $t$-dependence of the differential cross section was analyzed in detail and the total cross section for different combinations of vector mesons was calculated using the leading order impact factors and  BFKL amplitude. More recently, two other  studies on the process $\gamma^*\gamma^*\rightarrow VV$ have appeared in literature \cite{Enberg,DIvanov}. In the first one \cite{Enberg}, the leading order BFKL amplitude for the exclusive diffractive two-$\rho$ production in the forward direction is computed and the NLA corrections are estimated using a specific resummation of higher order effects.  In the last paper \cite{DIvanov}, the amplitude for the forward electroproduction of two light vector mesons in NLA is computed. In particular, the NLA amplitude is constructed by the convolution of the $\gamma^*\rightarrow V$ impact factor and the BFKL Green's function in the $\overline{\mathrm{MS}}$ scheme. In addition, a procedure to get results independent from the energy and renormalization scales has been investigated within NLA approximation.  A shortcoming of  those approaches is that they consider only the linear regime of the QCD dynamics and  nonlinear effects associated to the saturation physics are disregarded. However, double meson production in two photon interactions at high energies offers an ideal opportunity for studying the transition between the linear and saturation regimes since virtualities of both photons in the initial state can vary as well as the vector mesons in the final state. In the interaction of two highly virtual photons and/or double heavy vector meson production we expect the dominance of hard physics (linear regime). On the opposite case, characterized by double light vector meson production on real photons scattering, the soft physics is expected to be dominant. Consequently, for an intermediate scenario we may expect that the main contribution comes from semi-hard physics, determined by saturation effects.

In this paper we derive the main formulas to describe the double meson production in the dipole picture and  analyze the double  meson production in two photon interactions. We consider  three cases of physical and phenomenological interest: (a) the interaction of real photons, and the interaction of  virtual photons with (b) equal  and (c) different virtualities. In all cases we calculate the forward differential cross section for the $\rho \rho$, $\rho J/\Psi$ and $J/\Psi J/\Psi$ production. Moreover, we present a comparison between the linear and nonlinear predictions and estimate the  contribution for distinct photon polarizations.

\section{Basic Formulas}

\subsection{Double meson production in the dipole picture}
 
Let us introduce the main formulas concerning the vector meson
production in the color dipole picture. First, we consider the
scattering process $\gamma \gamma \rightarrow V_1 \, V_2$, where $V_i$ stands for
both light and heavy mesons. At high energies, the scattering process can be seen
 as a succession on time of three
factorizable subprocesses: i) the photon fluctuates in 
quark-antiquark pairs (the dipoles), ii) these color dipoles interact and, iii) the pairs convert into the vector mesons final states.
Using as kinematic variables the $\gamma^* \gamma^*$ c.m.s. energy
squared $s=W^2=(p+q)^2$, where $p$ and $q$ are the 
photon momenta,  the photon virtualities squared are given by 
$Q_1^2=-q^2$ and $Q_2^2 = -p^2$. The $x_{12}$ variable is defined by
\begin{eqnarray}
x_{12}= \frac{Q_1^2 + Q_2^2 + M_{V_1}^2 +  M_{V_2}^2}{W^2 + Q_1^2 + Q_2^2} \,\,.
\label{bjorken}
\end{eqnarray}
The
corresponding imaginary part of the amplitude at zero momentum
transfer reads as
\begin{eqnarray}
{\cal I}m \, {\cal A}\, (\gamma^* \gamma^* \rightarrow V_1 \, V_2) & = & \sum_{h, \bar{h}} \sum_{n, \bar{n}} 
\int dz_1\, d^2\rr_1 \,\Psi^\gamma_{h, \bar{h}}(z,\,\rr_1,\,Q_1^2)\,\, \Psi^{V_1*}_{h, \bar{h}}(z_1,\,\rr_1) \nonumber \\
&\times & \int dz_2\, d^2\rr_2 \,\Psi^\gamma_{n, \bar{n}}(z_2,\,\rr_2,\,Q_2^2)\,\, \Psi^{V_2 *}_{n, \bar{n}}(z_2,\,\rr_2)
\,
\sigma_{d d}(x_{12},\rr_1, \rr_2)
 \, ,
\label{sigmatot}
\end{eqnarray}
where $\Psi^{\gamma}$ and $\Psi^{V_i}$  are the light-cone wavefunctions  of the photon
  and vector meson, respectively. The quark and antiquark helicities are labeled by $h$, $\bar{h}$, $n$ and  $\bar{n}$
  and reference to the meson and photon helicities are implicitly understood. The variable $\rr_1$ defines the relative transverse
separation of the pair (dipole) and $z_1$ $(1-z_1)$ is the
longitudinal momentum fractions of the quark (antiquark). Similar definitions are valid for $\rr_2$ and  $z_2$. The basic
blocks are the photon wavefunction, $\Psi^{\gamma}$, the  meson
wavefunction, $\Psi_{T,\,L}^{V}$,  and the dipole-dipole  cross
section, $\sigma_{d\,d}$.

In the dipole formalism, the light-cone
 wavefunctions $\Psi_{h,\bar{h}}(z,\,\rr)$ in the mixed
 representation $(\rr,z)$ are obtained through two dimensional Fourier
 transform of the momentum space light-cone wavefunctions
 $\Psi_{h,\bar{h}}(z,\,\rk)$ (see more details, e.g. in Refs. \cite{predazzi,stasto,sandapen}). The
 normalized  light-cone wavefunctions for longitudinally ($L$) and
 transversely ($T$) polarized photons are given by:
\begin{eqnarray}
\Psi^{L}_{h,\bar{h}}(z,\,\rr)& = & \sqrt{\frac{N_{c}}{4\pi}}\,\delta_{h,-\bar{h}}\,e\,e_{f}\,2 z(1-z)\, Q \frac{K_{0}(\varepsilon r)}{2\pi}\,,
\label{wfL}\\
\Psi^{T(\gamma=\pm)}_{h,\bar{h}}(z,\,\rr) & = & \pm
\sqrt{\frac{N_{c}}{2\pi}} \,e\,e_{f}
 \left[i e^{ \pm i\theta_{r}} (z \delta_{h\pm,\bar{h}\mp} -
(1-z) \delta_{h\mp,\bar{h}\pm}) \partial_{r}
+  m_{f} \,\delta_{h\pm,\bar{h}\pm} \right]\frac{K_{0}(\varepsilon r)}{2\pi}\,,
\label{wfT}
\end{eqnarray}
where $\varepsilon^{2} = z(1-z)Q^{2} + m_{f}^{2}$. The quark mass
$m_f$ plays a role of a regulator when the photoproduction
regime is reached.  Namely, it prevents non-zero argument for the
modified Bessel functions $K_{0,1}(\varepsilon r)$ towards $Q^2\rightarrow 0$.
 The electric charge of the quark of flavor $f$ is given by $e\,e_f$.

For vector mesons, the light-cone wavefunctions are not known
in a systematic way and should be modeled. The
simplest approach  assumes a same vector  current as in the photon
case, but introducing  an additional vertex factor. Moreover, in
general the same functional form is chosen for the scalar part of the meson
light-cone wavefunction. Here, we follows the
analytically simple DGKP approach \cite{dgkp:97}. 
In this particular approach, one assumes
that the dependencies on $\rr$ and $z$ of the wavefunction are
factorised, with a Gaussian dependence on $\rr$.
Its main shortcoming  is that it breaks the rotational invariance between transverse and longitudinally polarized vector mesons \cite{Nikolaev}. However, as it describes reasonably the HERA data for vector meson production, as pointed out in Refs.
\cite{sandapen,magno_victor_mesons}, we will use it in our  phenomenological analysis. 
 The DGKP
longitudinal and transverse meson light-cone wavefunctions are
given by \cite{dgkp:97},
\begin{eqnarray}
\Psi_{h,\bar{h}}^{V,L}(z,\,\rr) & = &
z(1-z)\, \delta_{h,-\bar{h}}\,\frac{\sqrt{\pi} f_V}
{2\sqrt{N_{c}}\,\hat{e}_{f}}\,f_{L}(z)\,\exp
\left[\frac{-\,\omega_{L}^{2}\, \rr^{2}}{2} \right] \;,
\label{dgkp_L}\\
\Psi_{h,\bar{h}}^{V,T(\gamma = \pm)}(z,\,\rr) &=& \pm
\left(\frac{i\omega_T^{2}\,r e^{\pm i\theta_{r}}}{m_{V}}\,
[z\delta_{h\pm,\bar{h}\mp} -
(1-z)\delta_{h\mp,\bar{h}\pm}] + \frac{m_{f}}{m_{V}}\,\delta_{h\pm,\bar{h}\pm}
\right) \nonumber \\ & & \hspace*{4cm} \times
\frac{\sqrt{\pi} f_V}{\sqrt{2 N_{c}}
\,\hat{e}_{f}}f_{T}(z)\,\exp
\left[\frac{-\omega_{L}^{2} \rr^{2}}{2} \right] \;,
\label{dgkp_T}
\end{eqnarray}
where $\hat{e}_f$ is the effective charge arising from the sum
over quark flavors in the meson of mass $m_V$. The following
values $\hat{e}_f = 1/\sqrt{2}$ and $2/3$
stand for the $\rho$ and $J/\Psi$ mesons, respectively.
The coupling of the meson to electromagnetic current is labeled
by $f_V^2=3\,m_V\Gamma_{e^+e^-}/4\,\pi\alpha_{em}^2$ (see Table
\ref{tab:1}). The function $f_{T,\,L}(z)$ is given by the
Bauer-Stech-Wirbel model \cite{wsb:85}:
\begin{eqnarray}
f_{T,\,L}(z)= {\cal N}_{T,\,L}\,
\sqrt{z(1-z)}\,\exp \left[\frac{-\,m_{V}^{2}\,(z-1/2)^{2}}{2\,\omega^{2}_{T,\,L}}\right] \;.
\end{eqnarray}
The meson wavefunctions are constrained by the normalization
condition, which contains the hypothesis that the meson is composed only of
quark-antiquark pairs,  and  by the electronic decay width
$\Gamma_{V\rightarrow e^+e^-}$. Both
conditions are respectively  given by \cite{bl:80,stasto},
\begin{eqnarray}
& & \sum_{h,\bar{h}}\int d^{2}\rr \, dz  \,
|\Psi^{V(\lambda)}_{h,\bar{h}}(z,\,\rr)|^{2} = 1\,,
\label{norm1}\\
& & \sum_{h,\bar{h}}\int \frac{d^{2}\rr}{(2 \pi)^2}\, \frac{dz}
{z(1-z)}\, [z(1-z) Q^2 + k^2+m_f^2]\, \Psi^{V}_{h,\bar{h}}(k,z)
\Psi^{\gamma *}_{h,\bar{h}}(k,z)= e f_V m_V\, (\varepsilon_{\gamma}^{*}\cdot\varepsilon_{V})\,.
\label{norm2}
\end{eqnarray}
The constraints above, when used   on the DGKP wavefunction,
 imply  the following relations \cite{sandapen},
\begin{eqnarray}
& & \omega_{L,\,T} =   \frac{ \pi f_{V}}{\sqrt{2\, N_c} \hat{e}_{f}}
\,\sqrt{I_{L,\,T}}\,,
\label{dgkpnorm1}\\
& & \int_{0}^{1} dz \; z(1-z)\,f_{L}(z) = \int_{0}^{1} dz \,\frac{2\,[z^{2} +
    (1-z)^{2}]\,\omega_{T}^{2} + m_{f}^{2}}{2\,m_{V}^{2}\,z(1-z)}\,f_{T}(z) =1
\,,
\label{dgkpnorm2}
\end{eqnarray}
where
\begin{eqnarray}
I_{L} & = & \int_{0}^{1} dz \,z^{2}(1-z)^{2} \,f_{L}^{2}(z) \,,\\
I_{T} & = &   \int_{0}^{1} dz \,\frac{[z^{2} + (1-z)^{2}]\,\omega_{T}^{2} + m_{f}^{2}}{m_{V}^{2}}\, f_{T}^{2}(z) \,.
\end{eqnarray}
The relations in Eq. (\ref{dgkpnorm1}) come from the
normalization condition, whereas the relations in Eq.
(\ref{dgkpnorm2}) are a consequence of the leptonic decay width
constraints. The parameters $\omega_{T,\,L}$ and ${\cal
N}_{T,\,L}$ are determined by solving (\ref{dgkpnorm1}) and
(\ref{dgkpnorm2}) simultaneously. In Table \ref{tab:1} we quote the
results which will be used in our
further analysis. To be consistent with
the saturation models, which we will discuss
further, we have used the quark masses $m_{u,d,s}= 0.14$ GeV and
$m_{c}=1.5$ GeV.  We quote Refs. \cite{sandapen,magno_victor_mesons}
for more details in the present  approach and its comparison with data for
 both photo and electroproduction of vector mesons.

\begin{table}
\begin{center}
\begin{tabular}{||lcccccc||}
\hline
\hline
$V(m_V)$ & $\hat{e}_V$ & $f_V$ & $\omega_T$  & ${\cal N}_T$ & $\omega_L$  & ${\cal N}_L$  \\
 MeV &  & [GeV] & [GeV] &   & [GeV] &   \\
\hline
\hline
$\rho\,(770)$ & $1/\sqrt{2}$ & 0.153 & 0.218 & 8.682  & 0.331  & 15.091 \\
$J/\Psi\,(3097)$ & 2/3 & 0.270 & 0.546 & 7.665 & 0.680 & 19.350 \\
\hline
\hline
\end{tabular}
\end{center}
\caption{Parameters and normalization
  of the  DGKP vector meson light-cone wavefunctions. Results obtained using quark mass values from the saturation model (see text). }
\label{tab:1}
\end{table}

 Finally, the imaginary part of the forward amplitude can be obtained by
 putting the expressions for photon and vector meson (DGKP) wavefunctions,
 Eqs. (\ref{wfL}-\ref{wfT}) and (\ref{dgkp_L}-\ref{dgkp_T}), into
 Eq. (\ref{sigmatot}). Moreover, summation over the quark/antiquark
 helicities  and an average over the   transverse polarization states
 of the photon should be taken into account. 
In order to obtain the total cross section, we assume an exponential parameterization for the small $|t|$
behavior of the amplitude. After integration over $|t|$, the total
cross section for double vector meson production by real/virtual photons reads as,
\begin{eqnarray}
\sigma\, (\gamma \gamma \rightarrow V_1 \, V_2) = \frac{1}{B_{V_1 \,V_2}} \left. \frac{d\sigma  (\gamma \gamma \rightarrow V_1 \, V_2)}{dt}\,\right|_{t_{min}=0} =  \frac{[{\cal I}m \, {\cal A}(s,\,t=0)]^2}{16\pi\,B_{V_1 \,V_2}}\,(1+\beta^{2}) \;,
\label{totalcs}
\end{eqnarray}
where $\beta$ is the ratio of real to imaginary part of the
amplitude and $B_{V_1 \, V_2}$ is the slope parameter.

\subsection{Dipole-dipole cross section in the saturation model}
\label{dipdip}

The dipole formulation has been extensively used in the description of inclusive and diffractive processes at HERA  in an unified way. The basic quantity is the dipole-proton  cross section
$\sigma_{dip}$, which contains 
 all
information about the target and the strong interaction physics. In general, the saturation models \cite{GBW,bgbk,kowtea,iancu_munier}  interpolate between the small and large dipole
configurations, providing color transparency behavior, $\sigma_{dip} \sim \rr^2$, at $\rr \ll 1/Q_{\mathrm{sat}}$  and constant behavior at
large dipole separations $\rr > 1/Q_{\mathrm{sat}}$.  The physical scale which characterizes the transition between  the  dilute and  saturated system is denoted  saturation scale,  $Q_{\mathrm{sat}}^2\propto x^{-\lambda}$, which is energy dependent. Along these lines, the phenomenological saturation model proposed by Golec-Biernat and Wusthoff (GBW) \cite{GBW} resembles the main features of the Glauber-Mueller resummation. Namely, the dipole cross section in the GBW model takes the eikonal-like form,
\begin{eqnarray}
\sigma_{dip}^{\mathrm{GBW}} (x, \,\rr)  =  \sigma_0 \, \left[\, 1- \exp
\left(-\frac{\,Q_{\mathrm{sat}}^2(x)\,\rr^2}{4} \right) \, \right]\,. \label{gbwdip}
\end{eqnarray}
Its phenomenological application has been successful in a wide class of processes with a photon probe.
 Although the  GBW model describes reasonably well the  HERA data, its functional form is only an approximation of the theoretical nonlinear QCD approaches \cite{BK,CGC}. The parameters of the model are $\sigma_0=29.12$ mb, light quark masses ($u,d,s$) $m_f=0.14$ GeV, charm mass $m_c=1.5GeV$. Moreover, the saturation scale is given by $Q_{\mathrm{sat}}=(x_0/x)^{\lambda/2}$, with the parameters $x_0=4.1\cdot 10^{-5}$ and $\lambda = 0.277$.
In Ref. \cite{iancu_munier} a parameterization for the dipole cross section
was constructed to smoothly interpolate between the  limiting behaviors analytically under control: the solution of the BFKL equation
for small dipole sizes, $\rr\ll 1/Q_{\mathrm{sat}}(x)$, and the Levin-Tuchin law  \cite{levin}
for larger ones, $\rr\gg 1/Q_{\mathrm{sat}}(x)$. A fit to the structure function $F_2(x,Q^2)$ was performed in the kinematical range of interest, showing that it is  not very sensitive to the details of the interpolation. The dipole cross section was parameterized as follows,
\begin{eqnarray}
\sigma_{dip}^{\mathrm{IIM}}\,(x,\rr) = \sigma_0\, \left\{ \begin{array}{ll} 
{\mathcal N}_0\, \left(\frac{\rr\, Q_{\mathrm{sat}}}{2}\right)^{2\left(\gamma_{\mathrm{sat}} + \frac{\ln (2/\rr Q_{\mathrm{sat}})}{\kappa \,\lambda \,Y}\right)}\,, & \mbox{for $\rr Q_{\mathrm{sat}}(x) \le 2$}\,,\\
 1 - \exp \left[ -a\,\ln^2\,(b\,\rr\, Q_{\mathrm{sat}}) \right]\,,  & \mbox{for $\rr Q_{\mathrm{sat}}(x)  > 2$}\,, 
\end{array} \right.
\label{CGCfit}
\end{eqnarray}
where the expression for $\rr Q_{\mathrm{sat}}(x)  > 2$  (saturation region)   has the correct functional
form, as obtained either by solving the Balitsky-Kovchegov (BK) equation \cite{BK},
or from the theory of the Color Glass Condensate (CGC) \cite{CGC}. Hereafter, we label the model above by IIM. The coefficients $a=1.8$ and $b=1.13$ are determined from the continuity conditions of the dipole cross section  at $\rr Q_{\mathrm{sat}}(x)=2$. The coefficients $\gamma_{\mathrm{sat}}= 0.63$ and $\kappa= 9.9$  are fixed from their LO BFKL values. In our further calculations it will be used the parameters $R_p=0.641$ fm, $\lambda=0.253$, $x_0=0.267\times 10^{-4}$ and ${\mathcal N}_0=0.7$, which give the best fit result.
It is important to emphasize that the GBW and IIM saturation models are suitable in the region below $x=0.01$ and the
large $x$ limit needs still a consistent  treatment.
 At $ep$ collisions the dipole-proton cross sections should be supplemented by
a threshold factor
$(1-x)^{n_{\mathrm{thres}}}$, with $n_{\mathrm{thres}}=5$, which is directly associated with the number of spectators at $x \approx 1$ ($n_{\mathrm{thres}} = 2 n_{\mathrm{spect}} -1 $).
\begin{figure}[t]
\begin{tabular}{cc}
{\psfig{figure=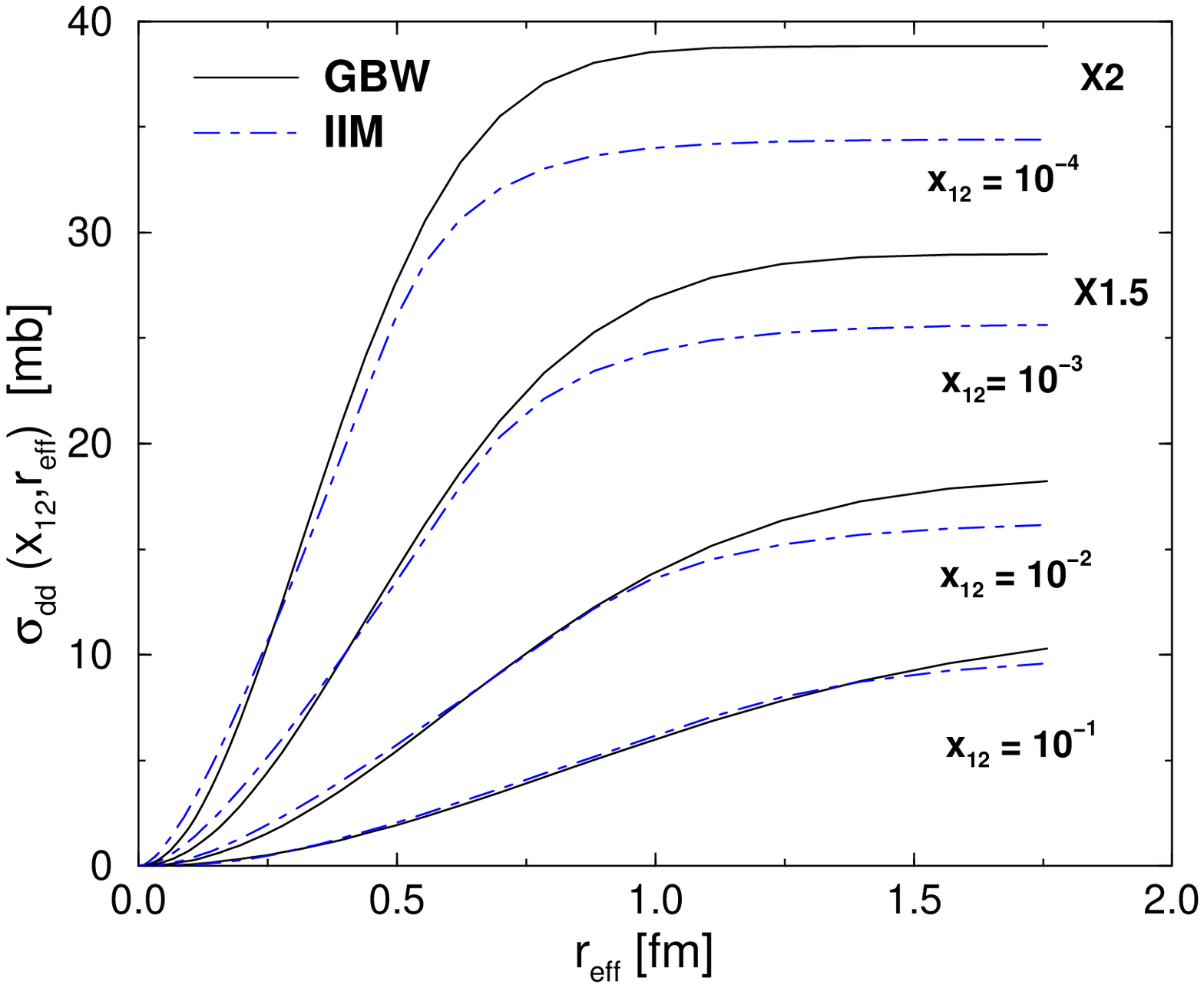,width=8.0cm}}  &
{\psfig{figure=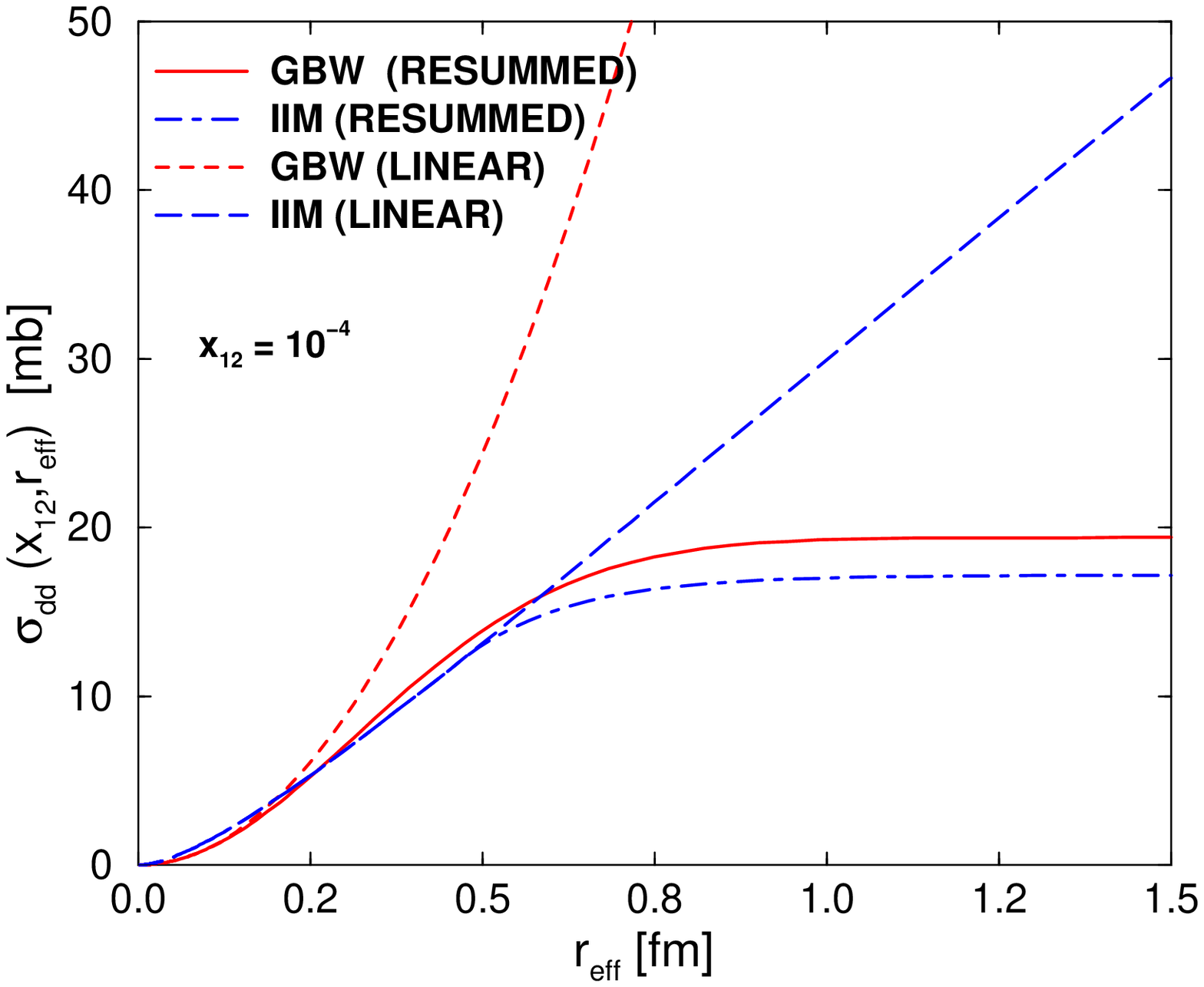,width=8.0cm}}\\
(a) &  (b)
\end{tabular}
\caption{Dipole-dipole cross sections: (a) Comparison between the predictions for  GBW and IIM models at different values of $x_{12}$. (b) Comparison between their resummed predictions and the corresponding linear limits.}
\label{fig1}
\end{figure}

Following Ref. \cite{Kwien_Motyka} we can extend  the saturation model, originally proposed to describe $ep$ collisions, to two-photon interactions at high energies. The basic idea is that the dipole-dipole cross section $\sigma_{dd} (x_{12},\rr_1,\rr_2)$  has the same functional form as the dipole-proton one and  is expressed in terms of an effective radius $\rr_{\mathrm{eff}}$, which depends on $\rr_1$ and/or $\rr_2$. Consequently, we have that \cite{Kwien_Motyka},
\begin{eqnarray}
\sigma_{dd}^{\mathrm{GBW}} (x_{12}, \,\rr_{\mathrm{eff}})  =  \hat{\sigma}_0 \, \left[\, 1- \exp
\left(-\frac{\,Q_{\mathrm{sat}}^2(x_{12})\,\rr^2_{\mathrm{eff}}}{4} \right) \, \right]\,, \label{gbwdippho}
\end{eqnarray}
and
\begin{eqnarray}
\sigma_{dd}^{\mathrm{IIM}}\,(x_{12},\rr_{\mathrm{eff}}) = \hat{\sigma}_0\, \left\{ \begin{array}{ll} 
{\mathcal N}_0\, \left(\frac{\rr_{\mathrm{eff}}\, Q_{\mathrm{sat}}}{2}\right)^{2\left(\gamma_{\mathrm{sat}} + \frac{\ln (2/\rr_{\mathrm{eff}} Q_{\mathrm{sat}})}{\kappa \,\lambda \,Y}\right)}\,, & \mbox{for $\rr_{\mathrm{eff}} Q_{\mathrm{sat}}(x_{12}) \le 2$}\,,\\
 1 - \exp \left[ -a\,\ln^2\,(b\,\rr_{\mathrm{eff}}\, Q_{\mathrm{sat}}) \right]\,,  & \mbox{for $\rr_{\mathrm{eff}} Q_{\mathrm{sat}}(x_{12})  > 2$}\,,
\end{array} \right.
\label{CGCfitpho}
\end{eqnarray}
where the  $x_{12}$ variable is given by the Eq. \ref{bjorken} and $\hat{\sigma}_0 = \frac{2}{3} \sigma_0$, with $\sigma_0$ the same as in Refs. \cite{GBW} and \cite{iancu_munier} and referred above.   The last relation can be justified in terms of the quark counting rule. In the two-photon case, the resulting quark masses are slightly different as it was found in Ref. \cite{Kwien_Motyka}: $m_f=0.21$ GeV for light quarks and $m_c=1.3$ GeV for charm. In Ref. \cite{Kwien_Motyka} three different scenarios for $\rr_{\mathrm{eff}}$ has been considered, with the dipole-dipole cross section presenting in all cases the color transparency property  ($\sigma_{dd} \rightarrow 0$ for $\rr_1 \rightarrow 0$ or $\rr_2 \rightarrow 0$) and saturation ($\sigma_{dd} \rightarrow \hat{\sigma}_0$) for large size dipoles. We quote also Ref. \cite{marquet}) for interesting discussions on the effective radius and its consequences in hadron-hadron interactions.  In what follows, we use the model I from \cite{Kwien_Motyka}, where  $\rr^2_{\mathrm{eff}} = \rr_1^2 \rr_2^2/(\rr_1^2 + \rr_2^2)$, which is favored by the $\gamma^* \gamma^*$ and $F_2^{\gamma}$ data.  We have tested the sensitivity of the result to a different prescription, $\rr^2_{\mathrm{eff}} = \mathrm{min}\,(\rr_1^2,\,\rr_2^2)$  (named model II in Ref. \cite{Kwien_Motyka}). Its deviation from  model I is quite large for $\rho$ production and  almost insensitive for the mixed $\rho J/\Psi$ production. For double $J/\Psi$ production the deviation is considerably larger than the mixed one. However, the difference concerns only to the overall normalization and no change is seen in the energy behavior. Moreover, in order to extend the dipole model to large $x_{12}$ it is necessary to take into account threshold correction factors which constrain that the cross section vanish when $x_{12} \rightarrow 1$ as a power of $1-x_{12}$. As in Ref. \cite{Kwien_Motyka}, we multiply the dipole-dipole cross section by the factor $(1-x_{12})^5$. A comment is in order here. One shortcoming of the GBW model is that it does not contain the correct DGLAP limit at large virtualities. Consequently, we may expect that its predictions are only valid at small values of the photon virtualities. Therefore, in what follows we only consider photon virtualities up to 10 GeV$^2$.

\begin{figure}[t]
\begin{tabular}{cc}
{\psfig{figure=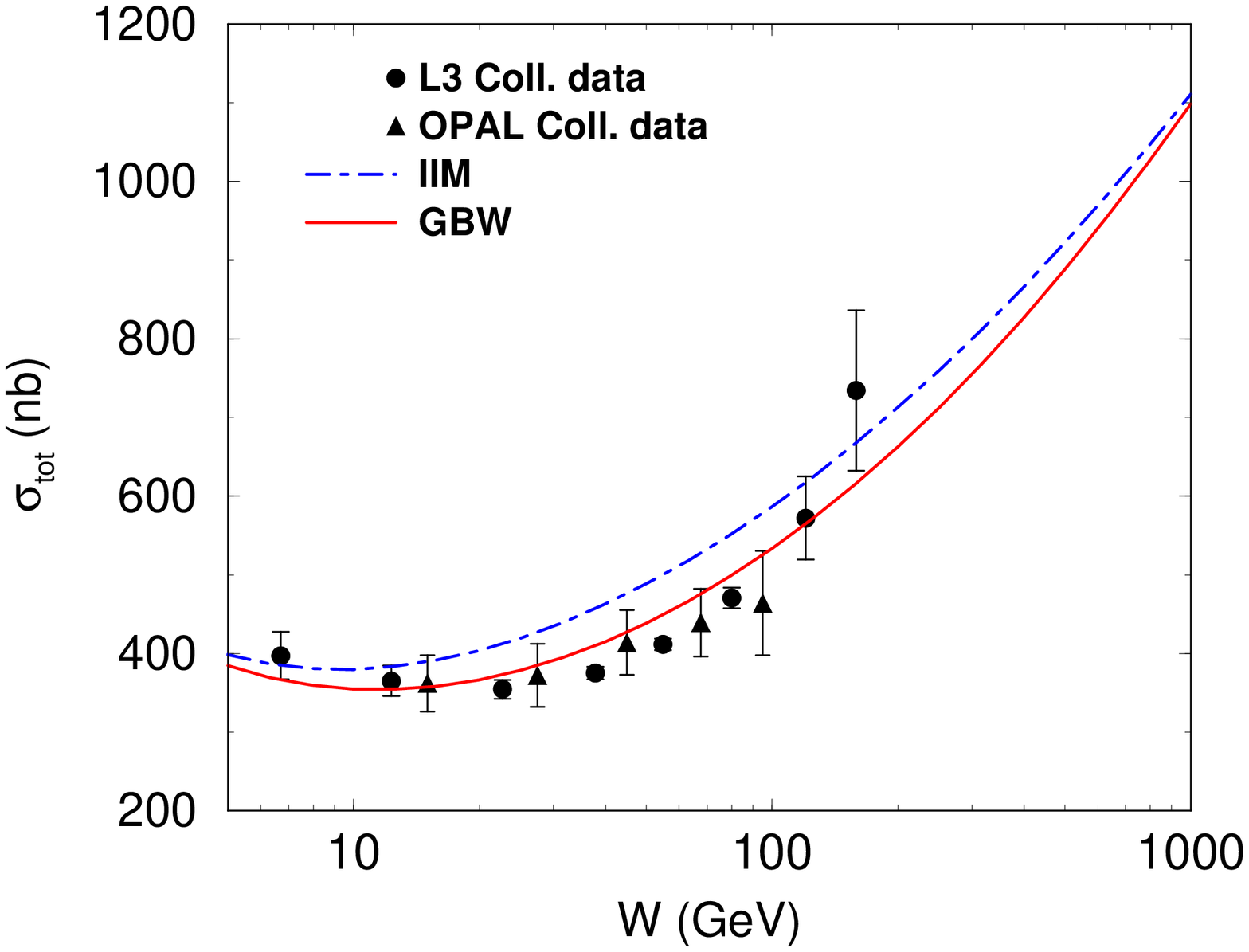,width=8.50cm}}  &
{\psfig{figure=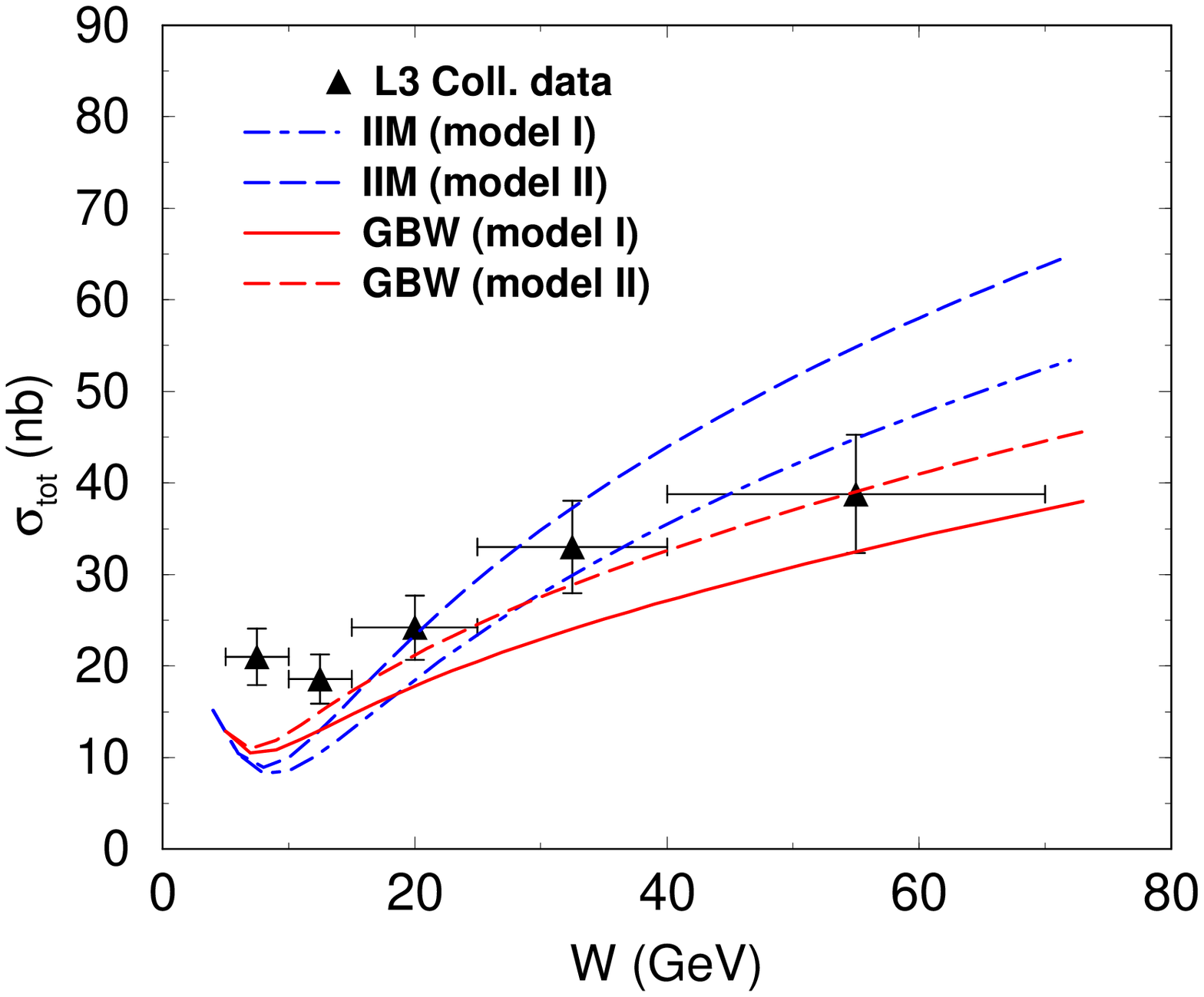,width=8.0cm}}\\
(a) &  (b)
\end{tabular}
\caption{A comparison between the predictions for (a) total cross section and (b) charm production in two real photon collisions considering the  GBW and IIM dipole-dipole cross sections. The QPM contribution has been added. Data from OPAL and  L3 Collaborations.}
\label{fig2}
\end{figure}

In Fig. \ref{fig1} (a) we present the dependence of the two dipole-dipole cross sections, Eqs. (\ref{gbwdippho}) and (\ref{CGCfitpho}), as a function the effective radius $\rr_{\mathrm{eff}}$ at different values of $x_{12}$ ($x_{12}=10^{-n},\,n=1,2,3,4$). We have that at small values of $\rr_{\mathrm{eff}}$ their  predictions  are similar, while they differ approximately 15 \% at large $\rr_{\mathrm{eff}}$ and small values of $x_{12}$. In order to emphasize the importance of the saturation effects, in Fig. \ref{fig1}-b we present a comparison between the full predictions of the GBW and IIM dipole-dipole cross sections and their linear limits. We have denoted  by resummed the curves with the complete expressions in Eqs. (\ref{gbwdippho}-\ref{CGCfitpho}) and by  linear their approximations  in the limit of small dipoles. Namely, for the linear case one has $\sigma_{dd}^{\mathrm{GBW}} \propto \hat{\sigma}_0(Q_{\mathrm{sat}}^2\rr^2_{\mathrm{eff}}/4)$ and for the IIM model we just take the extrapolation of Eq. (\ref{CGCfitpho}) for $\rr_{\mathrm{eff}} \leq 2/Q_{\mathrm{sat}}$.
We have that at $\rr_{\mathrm{eff}} \approx 0.2$ fm the linear and resummed predictions from the GBW model start to be different. On the other hand, in the IIM case, this difference starts at  $\rr_{\mathrm{eff}} \approx 0.5$ fm. Consequently, the transition between the linear and saturation regimes is distinct in the GBW and IIM models.



In next section we will compare the predictions for the double meson production coming from  different models for the dipole-dipole cross section. However, the extension of the IIM model for photon-photon interactions has not been considered before. For sake of completeness, we compare the predictions of the GBW and IIM  models for the specific cases of the total $\gamma \gamma$ cross section and  heavy  quark production. The analysis of the heavy quark production  is  motivated by its strict relation with the double $J/\Psi$ production. The GBW model has already been considered in Ref. \cite{Kwien_Motyka}, while the IIM analysis is  the first one  in literature. In Fig. \ref{fig2} (a) we present a comparison between the predictions of the GBW and IIM models  for the total cross section and the OPAL and L3 experimental data \cite{opal,l3}. Following \cite{Kwien_Motyka} we also include the QPM and reggeon contribution and assume the model I for the effective radius.  We have that the GBW and IIM  predictions  are similar, describing the current experimental data quite well.  It should be noticed that parameters for IIM have not been adjusted in order to fit two-photon data as done for GBW. Furthermore, we compute the cross section for charm production in the reaction $\gamma \gamma \rightarrow c\bar{c}$, considering real photons. The results are presented in Fig. \ref{fig2} (b) for two prescriptions of the effective radius (model I and II referred before) and compared with L3 data. The low energy quark box contribution (QPM) has been added. An additional contribution, which we do not include, is the resolved (single and double) piece to the charm cross section, which reaches 30 \% of the main contribution at high energies. As already verified in Ref. \cite{Kwien_Motyka}, both prescriptions for the effective radius provide  reasonable description of the data  when GBW model is considered. On the other hand, in the IIM model, the prescription I for the effective radius gives a better description, with the model II overestimating the L3 data at high energies. Moreover, the IIM model implies a stronger energy dependence of the heavy quark production cross section than the GBW prediction. This behavior should also be present in other processes characterized by a hard scale as, for instance, the interaction of two highly virtual photons or double heavy vector meson production. In what follows we only will consider the model I for the effective radius.


\section{Results}

In order to calculate the total cross section for double vector meson production given in Eq. (\ref{totalcs})  it is necessary to specify the value of the slope parameter $B_{V_1 \, V_2}$. As this quantity is not well constrained, in what follows we only will present our predictions  for the energy and virtuality dependence of  the forward differential cross section $ \left. \frac{d\sigma  (\gamma \gamma \rightarrow V_1 \, V_2)}{dt}\,\right|_{t_{min}=0}$. This should be enough for the present level of accuracy. We start by the study of the scattering of two real photons, investigating its dependence on energy and on the mesons mass. After, we consider the scattering of virtual photons and investigate the symmetric ($Q_1^2=Q_2^2$) and asymmetric   ($Q_1^2\propto \alpha Q_2^2$ with $\alpha \gg 1$) cases.  In addition, we estimate the magnitude of the contribution of the distinct polarizations for the total cross sections. Finally, we discuss the size of parton saturation effects in the production of different mesons.

\subsection{Double meson production on real photons interactions}

\begin{figure}[t]
\centerline{\psfig{figure=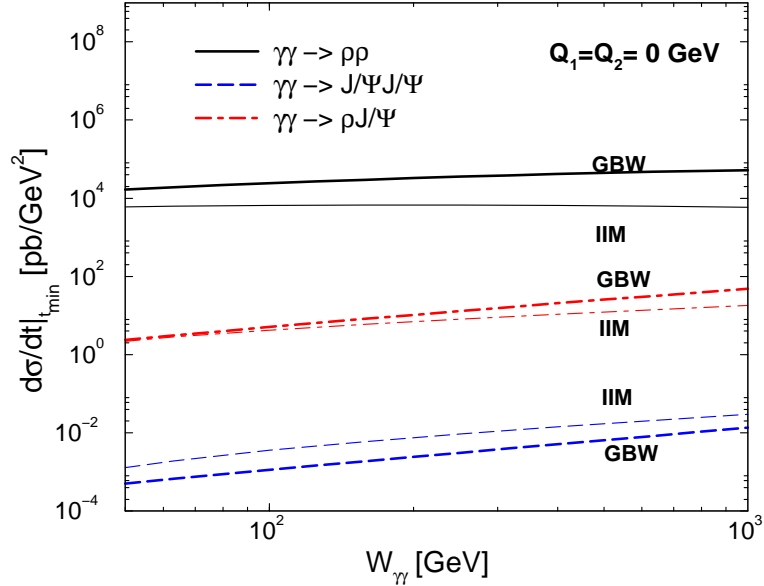,width=10cm}}
\caption{Energy dependence of the forward differential cross section  for double vector meson production considering real photons interactions ($Q_1^2=Q_2^2=0$ GeV$^2$). Bold (thin) curves are the results for GBW (IIM) parameterization for dipole-dipole cross section.}
\label{fig3}
\end{figure}

Let us start our analyzes considering the double meson production in two real photon scattering. In Fig. \ref{fig3} we present the forward differential cross sections for the representative cases of double $\rho $, $\rho J/\Psi$ and double $J/\Psi $ production in the energy range $50\, \mathrm{GeV}\leq W_{\gamma \gamma} \leq 10^{3}$ GeV. The curves are presented for the two models of the dipole-dipole cross section given in Eqs. (\ref{gbwdippho}) and (\ref{CGCfitpho}). Bold curves stand for GBW model and thin curves for IIM model.  The forward differential cross section is sizeable in the double $\rho$ case, being of order 20-40 nb/GeV$^2$ in the energy  range considered. The mixed $\rho J/\Psi$ production is the second higher rate reaching $4-40$ pb/GeV$^2$, whereas double $J/\Psi$ production is quite low. The deviations between the GBW and IIM  models  are large for the double $\rho$ production, with IIM results being a factor 10 below GBW one at $W_{\gamma \gamma} \approx 1$ TeV. The origin of this  discrepancy is not clear, since there is no evidence for strong deviations in the $\gamma \gamma $ case shown in Fig. \ref{fig2}-a. Probably, deviations could come from the different weights given by the wavefunctions in each case. This subject requires further investigations. For the double $J/\Psi$ production, the IIM prediction  overestimates the  GBW   by a factor 4, which agrees with the expectation which comes from our previous results for heavy quark production (See Fig. \ref{fig2}-b). On the other hand, in the $\rho J/\Psi$ case, the results are equivalent at low energies but differ by a factor two at 1 TeV, with the GBW prediction being greater than the IIM. These features can be qualitatively understood in terms of the scales involved in the process. As we discussed before, the IIM dipole-dipole cross section has a relatively faster transition to saturation in comparison with GBW and underestimates it by a factor of 20-30\% at small-$x_{12}$. In double $\rho$ and mixed vector meson production the typical scale is given by the light meson mass $\bar{\mu}^2=2M_{\rho}^2$ or the sum of light-heavy meson $\bar{\mu}^2=(M_{\rho}^2+M_{J/\Psi}^2)$. Therefore, the double $\rho$ process is dominated by a relatively soft scale and saturation effects should be important, whereas the mixed production is characterized by a semihard scale which is still sensitive to saturation effects. On the other hand, in the double $J/\Psi$ production the typical scale is sufficiently hard, $\bar{\mu}^2=2M_{J/\Psi}^2$. Therefore, we expect a larger contribution of small dipoles leading to a cross section with  higher magnitude.

In order to analyze the  energy dependence of the  forward differential cross section we have performed  a simple power-like fit in the energy interval $50\leq W_{\gamma \gamma}\leq 10^{3}$ GeV in the form $\left. \frac{d\sigma_{V_1V_2}}{dt}\,\right|_{t_{min}=0}\propto W_{\gamma \gamma}^{\alpha}$. For the double $\rho$ production one obtains  $\alpha=0.4 \,(0.08)$ for GBW (IIM)  model. In Regge phenomenology, this corresponds to an effective Pomeron intercept of order $\alpha_{\pom}^{\mathrm{eff}}\approx \alpha/4=0.1 \,(0.02)$ for GBW (IIM) parameterizations, which is clearly a soft behavior. This fact shows that the  IIM model contains stronger saturation effects in contrast to GBW one in the case of dominantly soft scales. In the mixed production, the effective power increases to $\alpha=0.96\,(0.65)$ for GBW (IIM) model and the difference  is not too sizeable as in the $\rho$ case. For the double $J/\Psi$ production, $\alpha=1.06\,(0.9)$, which implies  $\alpha_{\pom}^{\mathrm{eff}}\approx 0.27$. Therefore, one has a hard Pomeron behavior in the case  where the heavy meson mass ($\bar{\mu}^2=2M_{J/\Psi}^2$) is present in the problem. Thus, as expected from the phenomenology of $ep$ collisions, the saturation model for double vector meson production in $\gamma \gamma$ interactions is able to consistently  connect  the soft behavior when a non-perturbative scale is involved and the hard Pomeron  expectations when a perturbative scale is present.

\begin{figure}[t]
\begin{tabular}{cc}
\psfig{figure=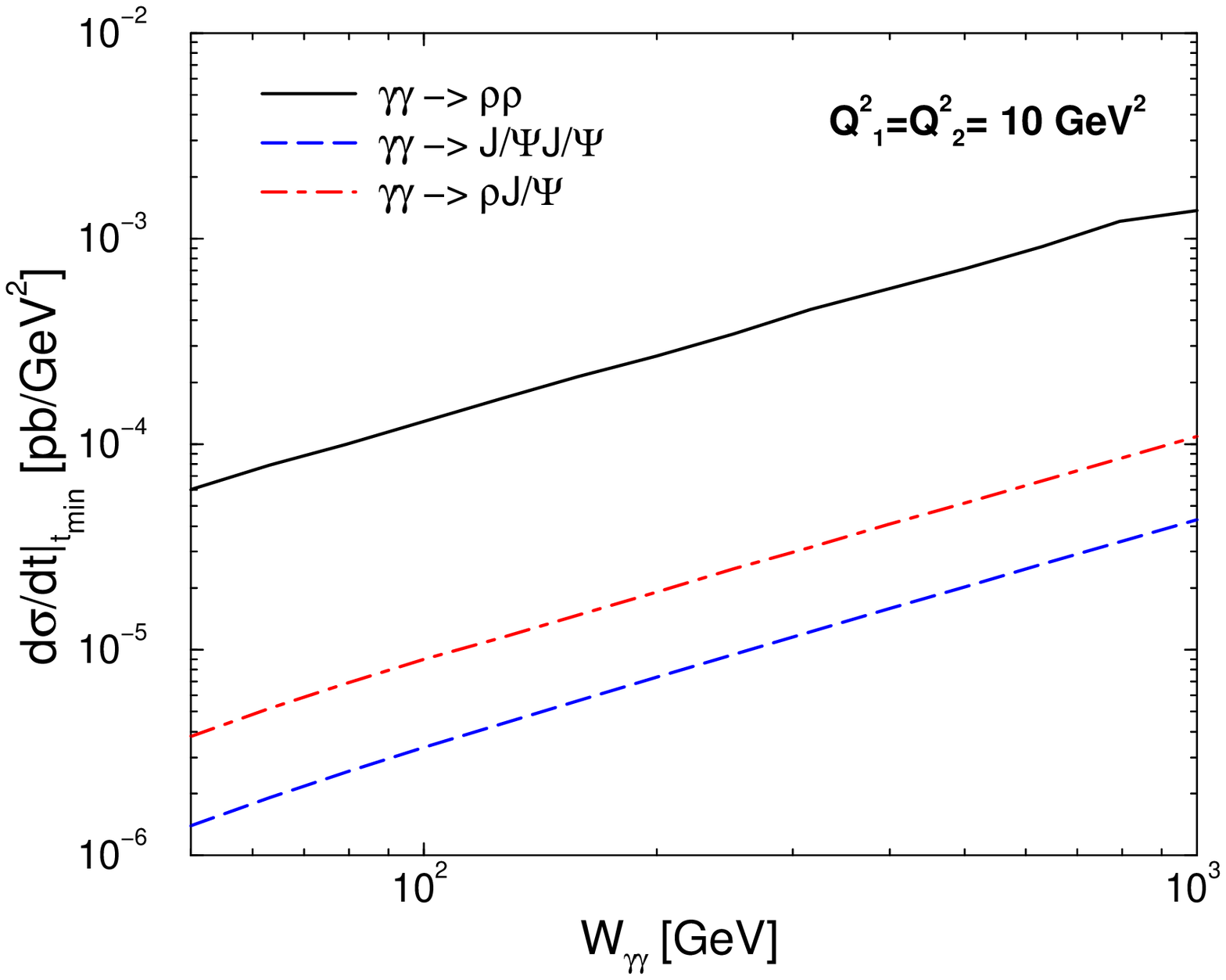,width=8.0cm}  &
\psfig{figure=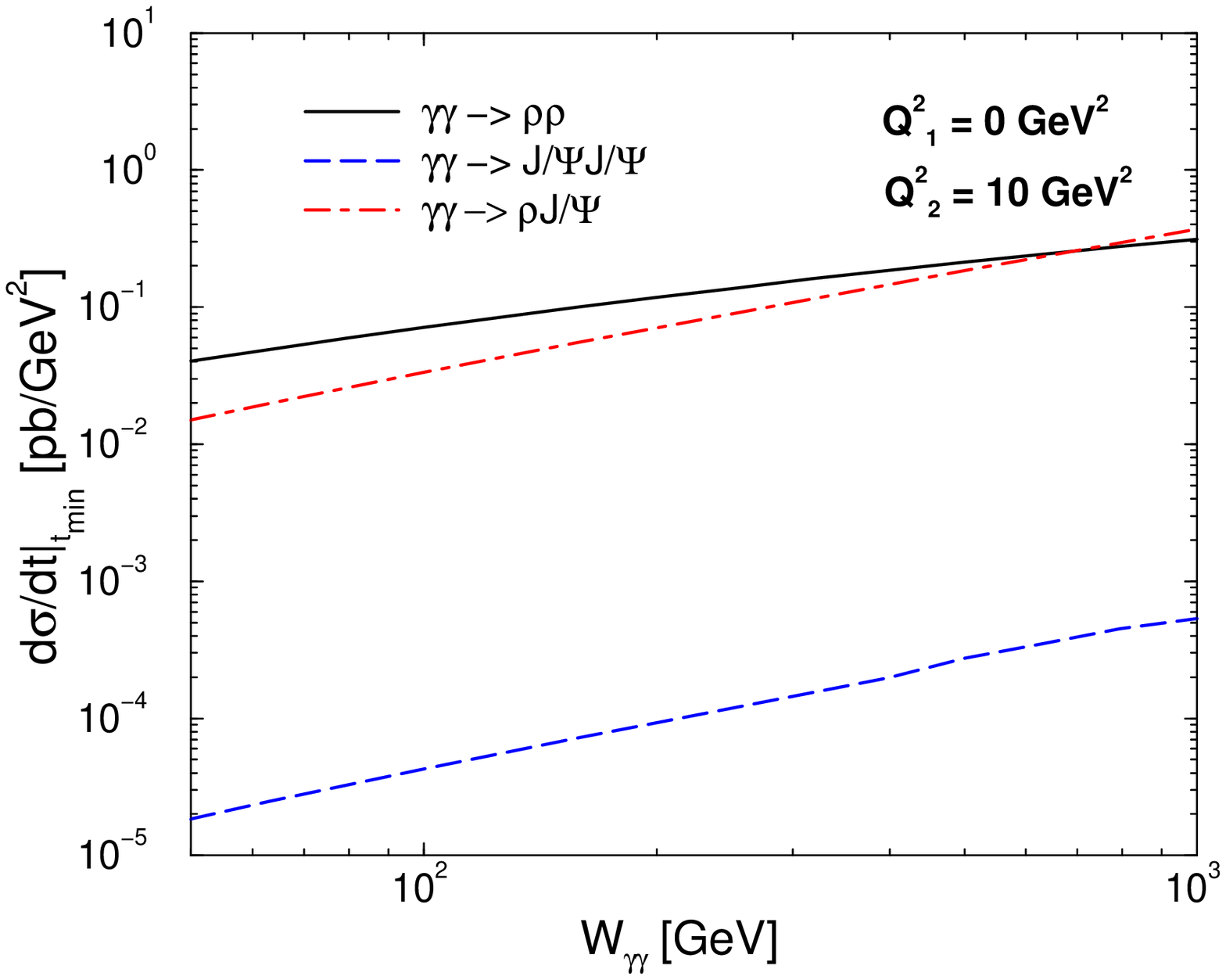,width=8.0cm}\\
(a) & (b)
\end{tabular}
\caption{Energy dependence for double vector meson production considering virtual photons considering: (a) equal virtualities ($Q_1^2=Q_2^2=10$ GeV$^2$) and (b) distinct virtualities ($Q_1^2=0$ and $Q_2^2=10$ GeV$^2$).}
\label{fig4}
\end{figure}

Let us now compare our results with those obtained in other approaches  \cite{dona_dosch,motyka,motyka_ziaja,per1,double_meson,vic_sauter}. Initially, let us consider the previous calculations within the color dipole picture \cite{dona_dosch}. In Ref. \cite{dona_dosch} there are estimations of the total cross section for double meson production. Our predictions underestimate those results by a factor ten for double $\rho$ and a factor one hundred for the other mesons. In this comparison we have used the following values for the slope parameters: $B_{\rho \rho}=10$ GeV$^{-2}$, $B_{\rho J/\Psi}=5$ GeV$^{-2}$ and $B_{J/\Psi J/\Psi}=0.44$ GeV$^{-2}$, which are taken from our recent investigations on double meson production in Refs. \cite{per1,double_meson}. These deviations are probably due to the different dipole-dipole cross section, distinct choices for the quark masses and uncertainties in the determination of the slope parameter. For instance, the dipole-dipole cross section in Ref. \cite{dona_dosch} behaves as $\sigma_{dd}\propto r_1^4r_2^4$ for small dipoles and $\sigma_{dd}\propto r_1^2r_2^2$ for large dipoles, which overestimate the integration on dipole sizes in comparison with the dipole-dipole cross sections presented here. Namely, one has $\sigma_{dd}\propto r_{\mathrm{eff}}^2Q_{\mathrm{sat}}^2$ for dipoles having transverse size $r_{\mathrm{eff}}<1/Q_{\mathrm{sat}}$ and $\sigma_{dd}\propto \hat{\sigma}_0$ for dipoles of size $r_{\mathrm{eff}}>1/Q_{\mathrm{sat}}$. Furthermore, the $\rho J/\Psi$ production in $\gamma \gamma$ processes has also  been estimated in Refs.\cite{motyka_ziaja,double_meson}. There,
the differential cross section was estimated in  a similar way as the elastic $J/\Psi$ photoproduction off the proton \cite{Ryskin}.  Our results agree with these predictions, with a behavior similar  to those obtained using the GRS(LO) parameterization for the gluon distribution on the  $\rho$ meson. This process was also estimated in Ref. \cite{vic_sauter} using the non-forward solution of the BFKL equation. Our results are smaller than estimations obtained in  Ref. \cite{vic_sauter}. This is expected since the saturation effects modifies significantly the  cross section of this semi-hard process.  Moreover, our results for the double $\rho$ production agree with those obtained in Ref. \cite{double_meson}  assuming the pomeron-exchange factorization.  However, as its predictions are strongly dependent on the assumptions present in the calculations of the double $J/\Psi$ and $\rho J/\Psi$ production (See Table 1 in Ref. \cite{double_meson}), a direct comparison is not very illuminating.  This process also was analyzed in Ref. \cite{vic_sauter}, but a direct comparison is not possible because only the hard contribution ($|t| > 1$ GeV$^2$) has been estimated.

\subsection{Double meson production on virtual photons interactions}

\begin{figure}
\centerline{\psfig{figure=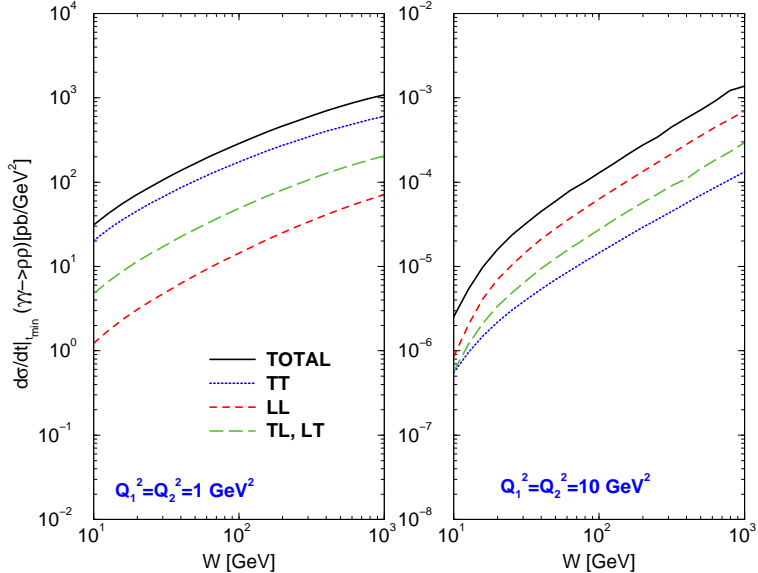,width=10cm}}
\caption{Energy dependence  for double $\rho$ production at equal photon virtualities, $Q^2=1$ and $Q^2=10$ GeV$^2$. The contributions of the different polarizations  (TT, LL, TL/LT) are explicitly presented. See discussion in text.}
\label{fig5}
\end{figure}

Let us now consider the double vector meson production when we have the interaction of virtual photons. In Fig. \ref{fig4} (a) we present the predictions of the GBW model  for the energy dependence considering that the incident photons have equal virtualities  ($Q^2=10$ GeV$^2$).  We have  that the forward  differential cross section decreases when the virtuality and/or the total mass of the final state is  increased. The differential cross sections  present a behavior similar on energy, independently of the meson mass. This is due the sufficiently hard scale for these processes given by $\mu^2=2Q^2+M_{V_1}^2+ M_{V_2}^2$, which is basically determined by the high photon virtuality, since $2Q^2 \geq M_V^2$. This also explains the proximity  between the  $\rho J/\Psi$ and double $J/\Psi$ predictions, in contrast with obtained for the real photon interactions.  A power-like fit to the differential cross section in the form $\left.\frac{d\sigma_{V_1V_2}}{dt}\,\right|_{t_{min}=0}\propto W_{\gamma \gamma}^{\alpha}$ gives $\alpha = 1.01, \,1.08,\,1.1$ for double $\rho$, $\rho J/\Psi$ and double $J/\Psi$, respectively. Our result for double $\rho$ is consistent with the NLA BFKL calculation using BLM scale fixing presented in Ref. \cite{Enberg}.

%

In Fig. \ref{fig4} (b) we present our predictions for double meson production considering  unequal photon virtualities. We consider the limit case of real photon scattering on a deeply virtual partner, namely $Q_1^2=0$ and $Q_2^2=10$ GeV$^2$. Now, the typical scale is given by $\mu^2= Q^2+M_{V_1}^2+ M_{V_2}^2$. In our computation of the mixed production we take the following statement for the photon virtualities: $Q_1^2$ corresponds to the photon transforming into $\rho$ and $Q_2^2$ corresponds to the photon transforming into $J/\Psi$. Notice that the final cross section should be given by $\sigma [\rho(Q_1)J/\Psi(Q_2) \,\mathrm{or}\, \rho(Q_2)J/\Psi(Q_1)]=  \sigma [\rho(Q_1)J/\Psi(Q_2)] + \sigma [\rho(Q_2)J/\Psi(Q_1)]$.  We have that the behavior of the different predictions are similar those obtained in Fig. \ref{fig4} (a), with the energy dependence for $\rho J/\Psi$ and double $J/\Psi$ production being almost identical those obtained in the symmetric case. The main difference occurs for double $\rho$ production, which has its energy dependence strongly modified by saturation effects due to the small value of $\mu^2$  present in the problem. A power-like fit to the differential cross section in the form $\left.\frac{d\sigma_{V_1V_2}}{dt}\,\right|_{t_{min}=0}\propto W_{\gamma \gamma}^{\alpha}$ gives $\alpha =0.65, \,1.03,\,1.08$ for double $\rho$, $\rho J/\Psi$ and double $J/\Psi$, respectively.

Using the dipole approach the contribution of the different polarizations for the forward differential cross section can be directly estimated. Let us start  considering the double $\rho$ production at equal virtualities of photons. We consider  the following illustrative cases, $Q_1^2=Q_2^2=1$ GeV$^2$ and  $Q_1^2=Q_2^2=10$ GeV$^2$. These choices allow to observe the dependence of each contribution on virtuality. The results are shown in Fig. \ref{fig5}. The transverse piece (TT) is labeled by dotted curves, longitudinal piece by dashed curves, mixed transverse-longitudinal (TL or LT) by long dashed curves and the total cross section (summation over polarizations) by the solid curves. In case of production of a same vector meson, the TL and LT pieces contribute equally, TL = LT. For virtualities $Q^2=1$ GeV$^2$, the transverse content dominates, followed by the LT/LT and LL pieces. Longitudinal content is a quite small contribution, which is consistent with the longitudinal wavefunction to be proportional to photon virtuality, which vanishes when $Q^2\rightarrow 0$. A completely different situation occurs  when the virtualities increase to $Q^2=10$ GeV$^2$. In this case  the longitudinal piece is dominant, followed by the LT/LT and transverse parts. This is consistent with the ratio $\sigma_L/\sigma_T\geq 0$ being $Q^2$-dependent in the light meson photoproduction (See e.g. Ref. \cite{Nikolaev}).

\begin{figure}
\centerline{\psfig{figure=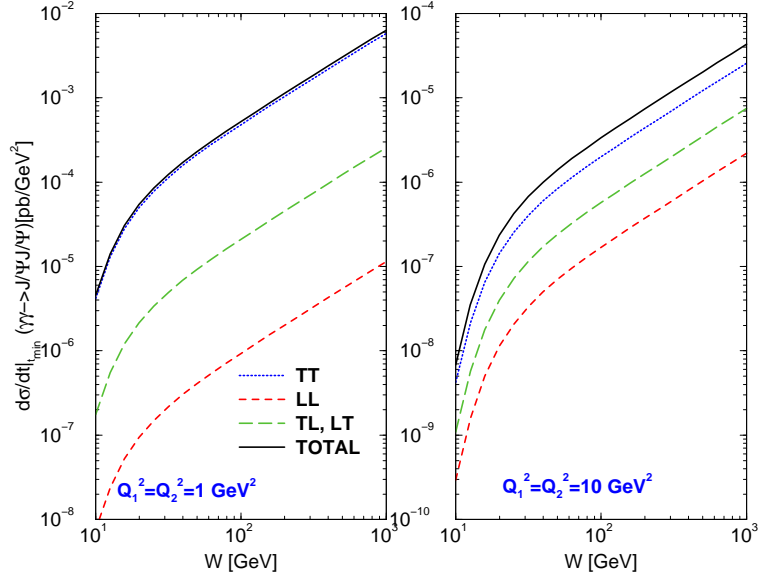,width=10cm}}
\caption{Energy dependence  for double $J/\Psi$ production at equal photon virtualities, $Q^2=1$ and $Q^2=10$ GeV$^2$. The contributions of the different polarizations  (TT, LL, TL/LT) are explicitly presented. See discussion in text.}
\label{fig6}
\end{figure}

A similar analysis can be made for the  double $J/\Psi$ production (see Fig. \ref{fig6}). We take the same virtualities for the virtual photons and equal notation as before. For virtualities $Q_1^2 = Q^2_2 = 1$ GeV$^2$, the transverse content dominates, followed by the LT/LT and LL pieces. As in the double $\rho$ case, the longitudinal contribution  is quite small. The total contribution is determined completely by the transverse contribution, with other pieces being negligible. A completely different situation occurs  when virtualities increase to $10$ GeV$^2$ in contrast with the $\rho$ case. The pattern remains the same as for $Q^2=1$ GeV$^2$, with transverse piece still dominant, followed by LT/TL and LL pieces. The total contribution is slightly larger than the transverse one.

\begin{figure}
\centerline{\psfig{figure=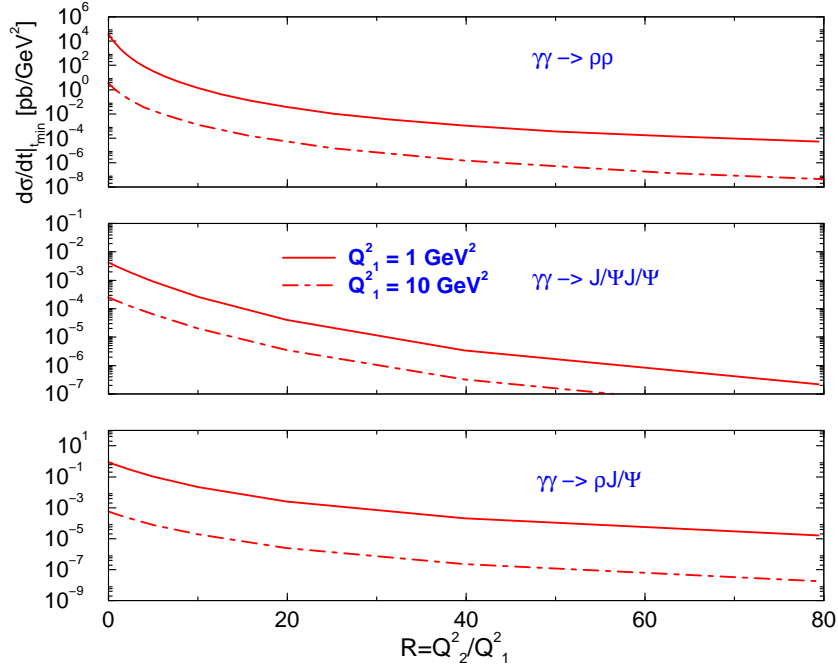,width=11cm}}
\caption{Dependence of the forward differential cross section  on the ratio $R=Q_2^2/Q_1^2$ of  photon virtualities at $Q_1^2$ fixed. }
\label{fig7}
\end{figure}

Finally, let us investigate the dependence on virtuality at fixed energy of the forward differential cross section. We take the representative energy of $W_{\gamma \gamma} = 500$ GeV. In Fig. \ref{fig7} we present the dependence of the forward differential cross section on the ratio $R=Q_2^2/Q_1^2$ at fixed $Q_1^2$. We consider the typical values $Q_1^2=1$ and $Q_1^2=10$ GeV$^2$. In all cases, the cross section decreases as $Q_2^2$ increases, presenting finite values towards  $R=0$. It should mentioned that this interpolation can not be obtained in the BFKL approach in view of the lack of a scale in the process. In our case, the saturation scale provides the semihard scale. In order to investigate the quantitative behavior on $R$ at intermediate virtualities, we adjust the curves with the simple exponential parameterization for $R\geq 1$ in the form $\left.\frac{d\sigma_{V_1V_2}}{dt}\,\right|_{t_{min}=0}\propto e^{-\beta R}$. This procedure gives $\beta = 4.4, \,2.6,\,2.9$ for double $\rho$, double $J/\Psi$ and $\rho J/\Psi$, respectively.  The results follow the  typical saddle-point  BFKL solution in the region of $R\geq 5$ GeV$^2$, namely the cross sections behave as $\left.\frac{d\sigma_{V_1V_2}}{dt}\,\right|_{t_{min}=0}\propto \frac{1}{Q_1^2Q_2^2}\exp (-\beta\ln^2R)$ as computed in Ref. \cite{Enberg}. It should be noticed that our definition for $R$ is slightly different from  that reference.

\subsection{Investigating saturation effects}

\begin{figure}
\centerline{\psfig{figure=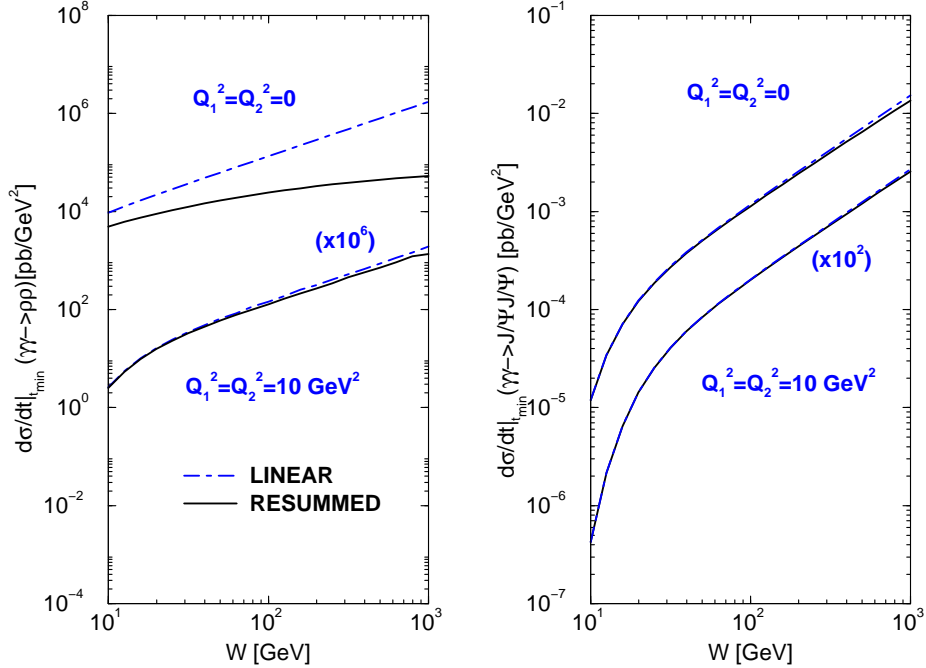,width=12cm}} 
\caption{Comparison between the linear and saturation predictions for the energy dependence of the forward differential cross section: (left panel) double $\rho$ production and (right panel) double $J/\Psi$ production. }
\label{fig8}
\end{figure}

Let us now investigate the magnitude of saturation effects in the differential cross section comparing the results using the small $r_{\mathrm{eff}}$ approximation for the dipole-dipole cross section  with the complete expression including the transition for  the  saturation regime (See discussion in Sect. \ref{dipdip}).  We restrict our analysis to a comparison between the linear and saturation model predictions for double $\rho$ and $J/\Psi$ production. The results are shown in Fig. \ref{fig8}, where solid lines stand for the resummed calculations and dot-dashed one for the linear approximation. It should be noticed that the saturation scale is different for each meson because $Q_{\mathrm{sat}}^2 \approx(x_0/x_{12})^{0.3}$ and $x_{12}$ depends on the meson mass as defined in Eq. (\ref{bjorken}). For the most striking case, in the production by two real photons, the saturation scale for $\rho$ reaches $Q_{\mathrm{sat}}^2 \approx 2.5$ GeV$^2$ whereas stays as $Q_{\mathrm{sat}}^2 \approx 1$ GeV$^2$ for $J/\Psi$. Therefore, the saturation scale is higher for $\rho$ than for $J/\Psi$ up to intermediate virtualities. Let us start discussing double $J/\Psi$ production [see Fig \ref{fig8} (b)], which is one typical hard process characterized by a hard scale given by $\mu^2(Q,M_V) = 2\,(Q^2 + M_{J/\Psi}^2)$. Consequently, we may expect that a perturbative description to be valid even in the real photon limit, $Q^2 \rightarrow 0$, and that the contribution from the saturation effects to be small while $\mu^2 \gg Q_{\mathrm{sat}}^2$. However, as the saturation scale grows with the energy, the saturation effects become important at large energies. This is the reason we observe a difference between linear and resummed predictions at $W\approx 1$ TeV in  the real photon case. These results  indicate that double $J/\Psi$ production is not strongly modified  by saturation corrections for energies smaller than 1 TeV. On the other hand, in the double $\rho$ production the situation changes drastically. For real photons, it is a typical soft process and, therefore, in this case the linear and saturation  predictions are very distinct. Now, the scale is given by $\mu^2(Q,M_V) = 2\,(Q^2 + M_{\rho}^2)$ , which should be treated carefully due to the small meson mass. The results are presented in Fig. \ref{fig8} (a). In the real photon scattering, one has $\mu^2=2M_{\rho}^2\approx 1$ GeV$^2$ and therefore saturation effects are increasingly important as $\mu^2 < Q_{\mathrm{sat}}^2(W_{\gamma\gamma})$, which explain the reason for the cross section to be reduced by two orders of magnitude at $W_{\gamma \gamma}\simeq 1$ TeV.  At  $Q^2=10$ GeV$^2$, the scenario is different since  $\mu^2=2Q^2 \geq Q_{\mathrm{sat}}^2$. Therefore, the resummed prediction is similar to the linear but corrections for large energies are still important. In view of discussions above, double $\rho$ production  becomes an ideal place to probe the saturation physics.

\section{Summary}

In this paper we have extended the dipole picture for the double vector meson production, $\gamma^* (Q_1) \gamma^*(Q_2) \rightarrow V_1V_2$, and calculated the forward differential cross section assuming that the dipole-dipole cross section can be modeled by a saturation model. We have analyzed the energy and virtuality dependence and investigated the magnitude of saturation effects. It is found that the effective power on energy is directly dependent on the typical momentum scale for the process, $\mu^2=Q_1^2 +Q_2^2+M_{V_1}^2+M_{V_2}^2$, which is different for distinct meson pair and photon virtualities. 
 Saturation effects are important for double $\rho$ production on real photons, whereas is small for processes containing $J/\Psi$ and/or large photon virtualities. It is shown the contribution of the distinct  polarizations and their regions of dominance for each meson pair. The results are consistent with expectations from  electroproduction of vector mesons. The dependence on virtuality has been investigated using the analysis on the ratio $R=Q_2^2/Q_1^2$. The results are qualitatively in agreement with previous predictions obtained using  dipole or NLO BFKL approaches. Our results demonstrate that 
double meson production in two photon interactions at high energies offer an ideal opportunity for the study of the transition between the linear and saturation regimes.

\section*{Acknowledgments}
VPG would like to thanks W. K. Sauter for  informative and  helpful discussions.  One of us (M. Machado) thanks the support of the High  Energy Physics Phenomenology Group, GFPAE IF-UFRGS, Brazil. The authors are grateful to C. Marquet for his valuable comments and suggestions.  This work was partially financed by the Brazilian funding agencies CNPq and FAPERGS.

\end{document}